# Tantalum electrodeposition using a nanoporous anodic alumina template and a nanostructured gold/nickel-chromium glass-ceramic substrate

Helena Simunkova[a,b*], Eva Kolibalova[b], Lukas Kalina[c], Tomáš Lednický[b,d], Jaromir Hubalek[a,b]

[a] *Department of Microelectronics, Brno University of Technology (BUT), Technicka 3058/10, Kralovo Pole, 61600 Brno, Czech Republic*

[b] *Central European Institute of Technology (CEITEC), BUT, Purkynova 123, 621 00 Brno-Medlanky, Czech Republic*

[c] *Faculty of Chemistry, Brno University of Technology (BUT), Purkynova 464/118, Brno 612 00, Czech Republic*

[d] *Leibniz Institute of Photonic Technology, Albert-Einstein-Str. 9, 07745 Jena, Germany*

*Corresponding author: simunkova@vutbr.cz

**Abstract**. Electrodeposition of tantalum coating was performed from an ionic liquid, BMP[Tf$_2$N], in the presence of dissolved anhydrous TaF$_5$ and LiF. Superficial X-ray photoelectron spectroscopy supplemented by an argon ion etching and depth profiling has proven the presence of the tantalum metal inside a thin coating deposited via a porous anodic alumina template. Additionally, tantalum electrodeposition was attempted using a "planar" sputter-deposited gold coating on a glass-ceramic substrate as the working electrode. Pores emerged within the sputter-deposited gold layer after the Ta electrodeposition step. Nano- to submicrometer large pores were created due to aluminum impurity diffusion through the gold and consequent etching effect of fluorides in the ionic liquid solution. Nanostructured tantalum and nanoporous gold are attractive materials with potential applications in high-performance electronic devices, including sensors, electrocatalysis, and energy storage systems.

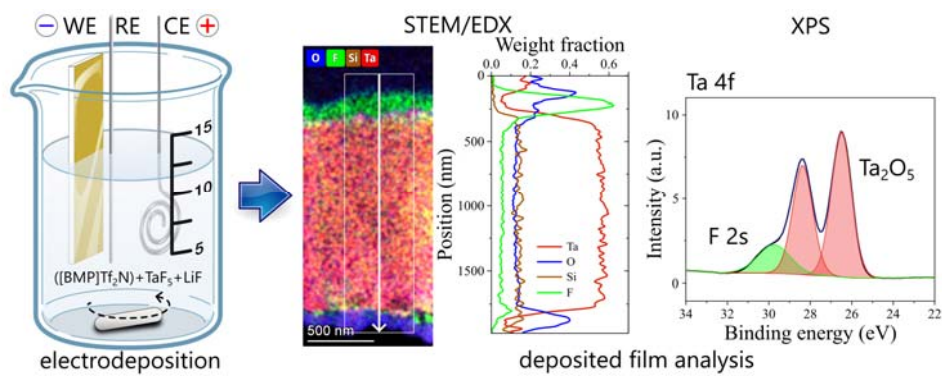

# 1. Introduction

Tantalum (Ta) is a typical refractory metal that is hard, ductile, biocompatible, and a good heat conductor [1, 2, 3, 4]. It is also known for its excellent resistance against acidic corrosion, which is partly due to a tantalum pentoxide ($Ta_2O_5$) passivation layer. Moreover, $Ta_2O_5$ is a promising sensing material with stable humidity and hydrogen sensitivity [5, 6], whereas the detection of ammonia was recently presented [7]. Tantalum cannot be electrodeposited from aqueous solutions. The reduction potential of tantalum is very low, making it difficult to reduce tantalum ions to metallic tantalum in an aqueous environment. Electrodeposition from ionic liquids and molten salts might be an alternative to traditional deposition techniques (PVD, CVD, and molten salt depositions). The molten salt systems are more conventional and better developed considering refractory metal electrodepositions than ionic liquid systems. Still, several studies of Ta electrodeposition from ionic liquids have already been published in detail [8, 9]. They achieved a thin crystalline Ta layer (up to 1 μm) deposited on specific planar substrates (such as platinum) [10]. Nahra *et al.* [1] studied the influence of working electrode material on the adherence and thickness of Ta electrodeposits. Gold-coated quartz single crystal, borosilicate glass, or mica substrates [10, 11, 12] were often applied as the working electrode materials during Ta electrodeposition from fluoride-containing ionic liquid solutions. However, no experimental study was provided on the silicon content inside the electrodeposited Ta coating. Herein, we present Ta electrodeposition using two diverse working electrodes (substrates): 1) a nanoporous templated electrode consisting of porous anodic alumina (PAA) superimposed over a metallic conductive layer deposited on a silicon wafer and 2) a "planar" gold/nickel-chromium (Au/NiCr)/glass-ceramic substrate. Our previous study described Ta electrodeposition via the PAA template with a conductive base, most probably for the first time [13]. Well-aligned and mechanically stable nanotubes were obtained after the PAA template dissolution. Herein, we performed detailed X-ray photoelectron spectroscopy (XPS) analyses supplemented by an argon ion (Ar$^+$) etching and depth profiling to prove Ta metal content within the nanotubes.

This work was undertaken to understand Ta electrodeposition from an ionic liquid solution proceeding at lower temperatures than the molten salt systems, which led to significant energy savings.

## 2. Material and methods

Tantalum (Ta) electrodeposition was attempted using a room temperature ionic liquid (IL) 1-butyl-1-methylpyrrolidinium bis(trifluoro-methylsulfonyl) imide, ([BMP]Tf$_2$N) (Solvionic, 99.9 %), containing TaF$_5$ (Alfa Aesar, 99.9 %) and LiF (Alfa Aesar, 99.99 %). The salts were dissolved in the IL and held for about 1 hour at 120°C to remove trace water. The entire setup was handled under inert gas (nitrogen) conditions (H$_2$O, O$_2$ < 15 ppm). A potentiostat (Autolab PGSTAT204/FRA32M, Metrohm) was used as the current source. A glass flask cell was covered with a PTFE lid, holding the working and reference electrodes. A platinum (Pt) wire was used as a counter electrode. Two diverse working electrodes were applied, including a porous anodic alumina (PAA) template with a conductive base (valve metal); see ref. [14] and secondly, a planar glass-ceramic Sitall substrate, covered with polycrystalline gold (Au) (ca. 300 nm thick, on a 20 nm thick nickel-chromium NiCr layer as an adhesion promoter, made by magnetron sputtering), designated as Au/NiCr/glass-ceramic. Ta was electrodeposited potentiostatically at −3 V vs. a platinum (Pt) wire pseudo reference electrode for 10 seconds or potentiodynamically (in a range from −1.4 V to −1.7 V and back to −1.4 V vs. Pt), in a glove box under inert gas N$_2$, at 200 °C. Both TaF$_5$ and LiF were either 0.25 M or 0.4 M, respectively. The PAA template was etched in an aqueous phosphoric acid and chromium trioxide solution.

Transmission electron microscopy (**TEM**) and scanning transmission electron microscopy (**STEM**) using Thermo Fisher Scientific Titan Themis 60–300 (scanning) transmission electron microscope operated at 300 kV were utilized to investigate the coating cross-section prepared in Focused ion beam (**FIB**)/scanning electron microscope FEI Helios NanoLab 660. Energy-dispersive X-ray spectroscopy (**EDX**) was used to determine the atomic composition of the coated layers. The EDX analyses were conducted at the acceleration voltage of 300 kV.

X-ray photoelectron spectroscopy (**XPS**) analyses were carried out with an Axis Ultra DLD spectrometer using a monochromatic Al Kα (hν = 1486.7 eV) X-ray source operating at 150 W (10 mA, 15 kV). The spectra were obtained using an analysis area of ~300 × 700 μm$^2$. The Kratos charge neutralizer system was used for all analyses. High-resolution spectra were measured with a step size of 0.1 eV and 20 eV pass energy. The instrument base pressure was 2·10$^{-8}$ Pa. Spectra were analyzed using CasaXPS software (version 2.3.15) and have been charged corrected to the main line of the carbon C 1s spectral component (C–C, C–H) set to 284.80 eV. A standard Shirley background was used for all sample spectra. Ar$^+$ etching removed approximately 18–64 nm of the superficial material. The analysis was performed on Ta electrodeposited via the PAA template and on Ta coating deposited on the "planar" Au/NiCr/glass-ceramic working electrode.

## 3. Results and discussion

*3.1. Tantalum electrodeposition via porous anodic alumina template*

Herein, tantalum (Ta) nanotubes were electrodeposited potentiostatically (−3 V vs. Pt) in an air- and water-stable ionic liquid: 1-butyl-1-methylpyrrolidinium bis(tri-fluoromethylsulphonyl) imide ([BMP]Tf$_2$N), using an anhydrous salt, TaF$_5$ (0.25 M) as the source of Ta, together with LiF (0.25 M). A template made of a porous anodic alumina (PAA) was employed. One face of the PAA template was rendered conductive to enable the electrodeposition. More on the preparation of the PAA template was published elsewhere [ 14 ]. A general schematic diagram of the Ta electrodeposition through the PAA upon an underlying metal is shown in Figure 1. In our previous paper [ 15 ], Ta nanotube arrays were electrodeposited and analyzed. Numerous chronoamperometry and charge density analyses revealed that tantalum metal was the most probable component of the electroreduction via the PAA template. Nevertheless, due to the high Ta metal affinity for oxygen, $Ta_2O_5$ was mainly detected upon its surface.

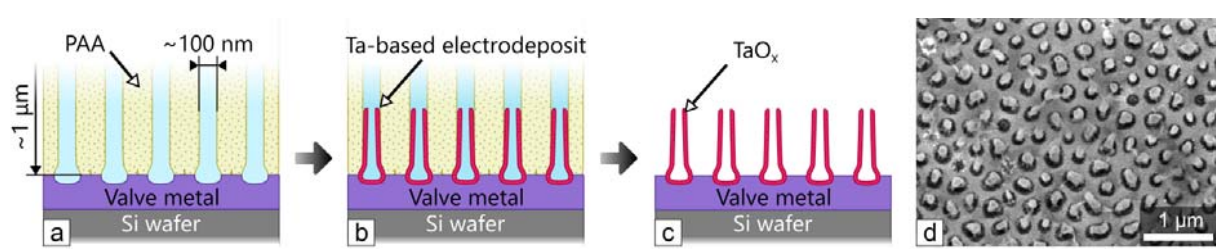

*Figure 1 a)–b) Schematic of tantalum electrodeposition through a PAA template upon an underlying valve metal followed by c) selective removal of PAA. The image d) shows tantalum nanotube arrays (illustrated in c).*

Herein, an additional XPS analysis supplemented by an $Ar^+$ etching of the free-standing nanotubes was performed, verifying the tantalum metal content inside the nanotubes. No Ta metal was detected before the $Ar^+$ etching. The surface analysis revealed only $Ta_2O_5$. After etching, which removed about 18 nm of the electrodeposited material, 65 at. % of Ta metal was revealed (green curve in Figure 2), along with about 5 at. % of Ta suboxide (red curve in Figure 2) according to the Ta4f spectrum. The remaining material was $Ta_2O_5$ (about 30 at. %). The Ta metal content increased up to 69 at. % after removing more than 60 nm of the material. The initial surface's survey and high-resolution (Ta4f) XPS spectra are shown at the top, while the high-resolution depth profiles are shown at the bottom of Figure 2. The surface contained some impurities, such as carbon, phosphorus, and a negligible amount of fluoride. Phosphorus originated from the PAA template etchant, while carbon and fluorides were part of the electrodeposition solution. These XPS analyses support our previous findings based on charge density analysis, which determined that the main component of the electroreduction is tantalum metal [15].

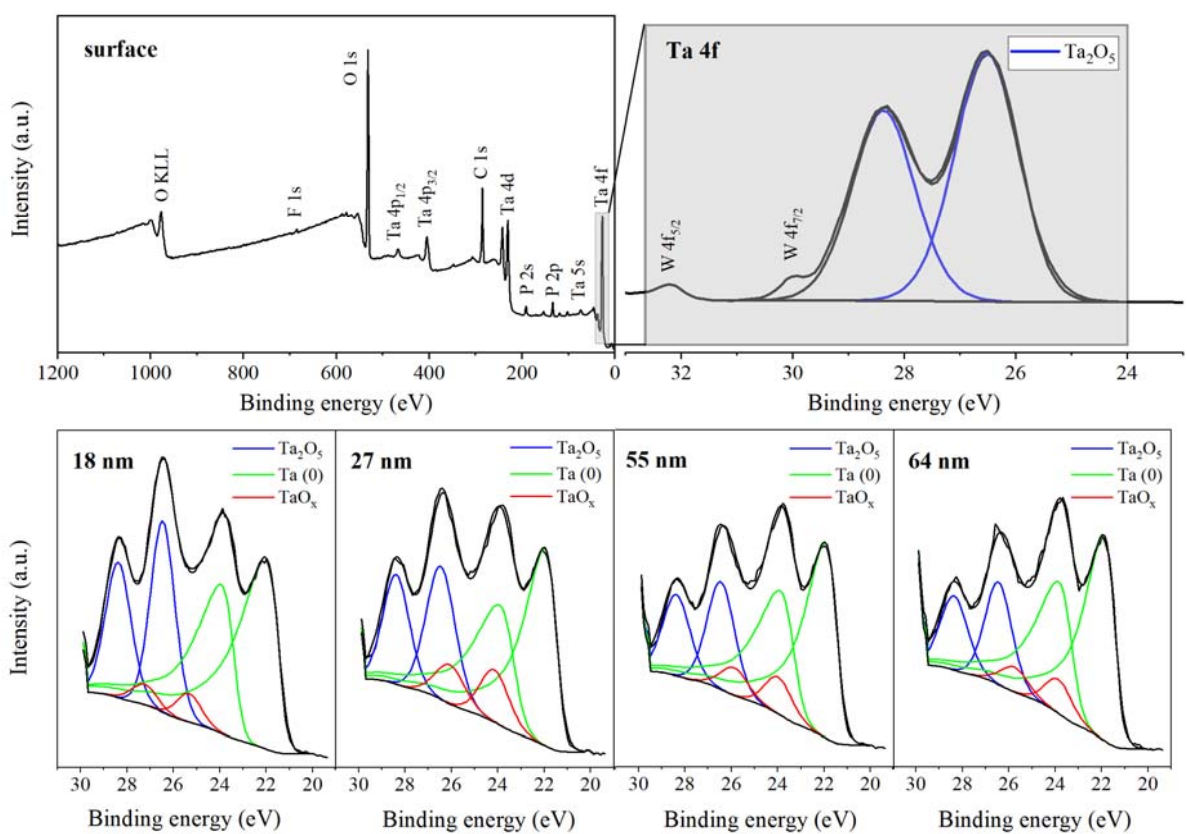

*Figure 2 XPS analysis of tantalum nanotube arrays: (at the top) shows the surface analysis, (at the bottom) shows the depth profiling by Ar+ etching (labeled by the amount of removed material).*

*3.2. Tantalum electrodeposition via nanoporous gold substrate*

Tantalum electrodeposition was further performed using a "planar" working electrode made of glass-ceramic (Sitall), covered with 20 nm nickel-chromium (NiCr) adhesive coating, and 300 nm sputter-deposited gold (Au) on the top. The word "planar" (in parentheses) means cavities that emerged within the Au layer during the Ta electrodeposition process. The experimental conditions were the same as in the case of the PAA template. The choice of the glass-ceramic working substrate resulted in oxygen, silicon, and aluminum impurities within the Ta coating. The fluoride ions in the ionic liquid solution caused the etching effect. Ta metal was not determined in any specimen using that substrate (with or without a postprocedural annealing treatment). Herein, XPS was performed after annealing at 600 °C for 4 hours at a pressure of approximately $10^{-4}$ Pa, supplemented by the Ar+ etching to remove 20 nm of the surface material (Figure 3). The presence of metal Ta was not observed even after removing the superficial part of the coating. Ta was found mainly as pentoxide $Ta_2O_5$ and a minor part of suboxide $TaO_x$ (6 at. %) upon the very surface. The $Ar^+$ etching resulted in 23 at. % tantalum carbide ($TaC_x$) and 12 at. % $TaO_x$, and the rest of $Ta_2O_5$. In our previous study [15], it was determined that carbon partially turns into $TaC_x$ due to a post-deposition annealing step. Herein, the presence of $TaC_x$ inside the Ta coating was additionally proved, indicating that carbon-containing ionic liquid was trapped within that coated layer. Moreover, the surface also contained chromium, which originated from the NiCr adhesive layer.

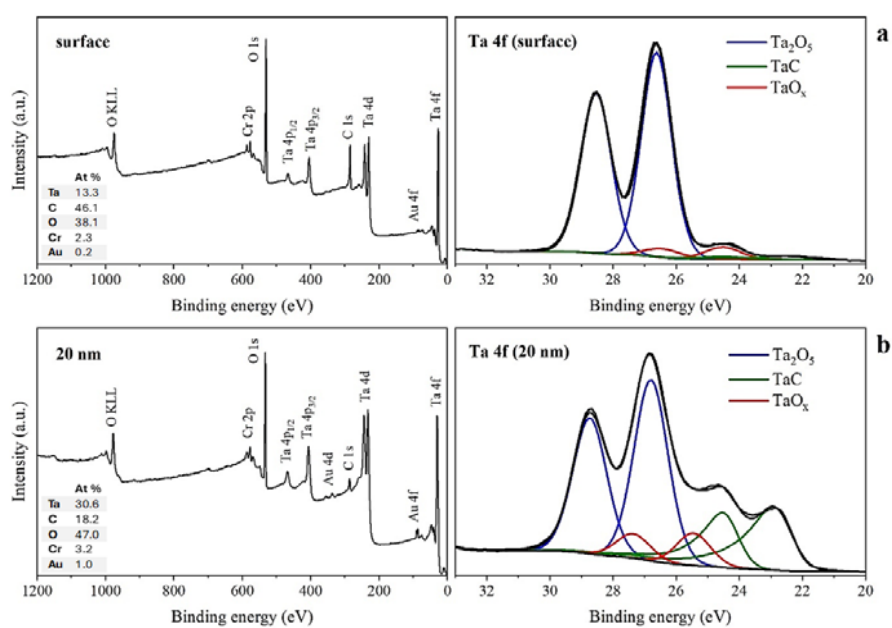

*Figure 3 XPS analyses of tantalum electrodeposited coating on Au/NiCr WE, both survey and high-resolution spectra of (a) surface and (b) depth profile after 20 nm $Ar^+$ etching. The sample was annealed at 600 °C for 4 hours at a pressure of approximately $10^{-4}$ Pa.*

A STEM-EDX experiment was conducted to analyze the element distribution in both the electrodeposited Ta and the Au sputter-deposited underlayer. The specimen was electrodeposited at a higher fluoride concentration (0.4 M of both $TaF_5$ and LiF) if compared to the previous XPS analysis. The thickness of the Ta coating was approximately 1,430 nm. Both oxides and fluorides were mainly accumulated at the top and bottom of the Ta coating, as shown in Figure 4. The bottom impurity layer, approximately 150 nm thick oxide (containing Si and Al), formed during a pre-heating step, which took about 43 seconds before the start of the Ta electrodeposition. The EDX analysis proved significant Si, O, F, C, and Al content in the Ta coating. See Table 1 for the atomic and weight elemental fractions and Figure 5 for the element distribution across the Ta coating. Considering the element distribution plot and the color EDX maps in Figure 4, the coating composition appears relatively uniform, indicating that the deposition was not significantly affected by changes in local ionic concentrations.

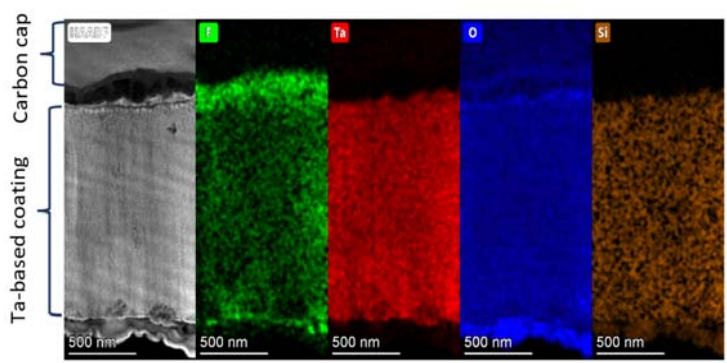

*Figure 4 HAADF - STEM image and corresponding STEM-EDX elemental mapping of tantalum-based coating, FIB cross-section upon gold / Sitall glass-ceramic substrate. Same specimen as in Figure 6. (Deposited by cycling from −1.4 V to −1.7 V and back to -1.4 V vs. Pt, at zero stirring rate from 0.4 M TaF5 and 0.4 M LiF in [BMP]Tf$_2$N ionic liquid solution).*

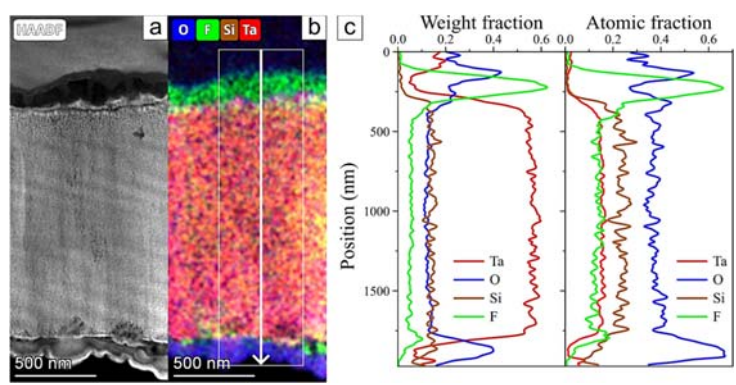

*Figure 5 Cross-section of tantalum-based coating on gold / Sitall glass-ceramic substrate (a) HAADF - STEM picture, (b) corresponding EDX color mix map (O-K, F-K, Si-K, Ta-L), (c) spectrum profile plot showing element distribution as weight and atomic fraction across tantalum-based coating. Same specimen as in Figure 4. (Electrodeposited by cycling from −1.4 V to −1.7 V and back to −1.4 V vs. Pt, at zero stirring rate from 0.4 M TaF5 and 0.4 M LiF in [BMP]Tf2N ionic liquid solution).*

*Table 1 Compositional fractions inside the tantalum-based coating, of which cross-section was shown in Figure 4 and Figure 5.*

| Element | Family | Atomic Fraction (%) | Atomic Error (%) | Mass Fraction (%) | Mass Error (%) | Fit Error (%) |
|---|---|---|---|---|---|---|
| C  | K | 10.6 | 1.9  | 3.4  | 0.6  | 17.4 |
| O  | K | 37.3 | 5.2  | 15.8 | 2.9  | 2.6  |
| F  | K | 15.3 | 3.0  | 7.7  | 1.5  | 3.3  |
| Al | K | 4.8  | 1.0  | 3.4  | 0.7  | 3.5  |
| Si | K | 20.4 | 3.6  | 15.2 | 2.7  | 0.8  |
| Ca | K | 0.3  | 0.04 | 0.3  | 0.04 | 1.7  |
| Ti | K | 0.1  | 0.02 | 0.1  | 0.02 | 0.8  |
| Ta | L | 11.1 | 1.5  | 53.2 | 3.7  | 0.3  |
| Au | L | 0.2  | 0.02 | 0.9  | 0.1  | 1.8  |

The black appearance of most of our deposits indicated the presence of oxidized compound(s) on top of the coatings. Black-colored coatings were determined by others too [16]. At the large potential limit (±3.5 V) experiment, the ionic liquid solution became black due to a powdery black deposit released into the solution. Similarly, Nahra *et al.* [11] observed that at potentials higher than −2.4 V (vs Pt-Fc/Fc+) a black powder was stripped from the working electrode surface to the electrolyte. An irreversible behavior was observed at a cyclic voltammetry potential range from open circuit potential OCP to −1.6 V and to +0.5 V (Figure 1 in Supplementary). It was noted elsewhere that oxides are very reactive with tantalum fluorides leading to a formation of tantalum oxyfluorides: $TaOF_5^{2-}$ and $TaO_2F_x^{(x-1)-}$ [17]. It is supposed that the irreversibility resulted from the choice of the working electrode material, glass-ceramic, which was attacked by fluoride ions and released oxide ions. These oxide ions could complex the tantalum ions, giving tantalum oxyfluorides, which might be reduced to some oxidized black-colored tantalum compound(s).

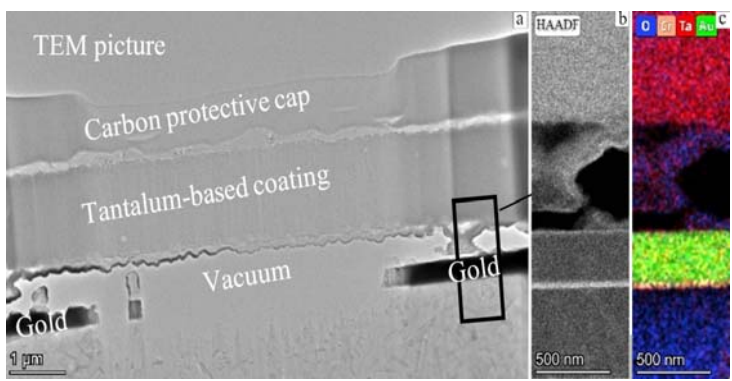

*Figure 6 Cross-section of tantalum-based coating on Au/NiCr on Sitall glass-ceramic substrate (a) TEM overview image, (b) HAADF - STEM (high-angle annular dark field scanning transmission) picture and (c) EDX color mix map (O-K, Cr-K, Ta-L, Au-L) origin from same sample cross-section. (Deposited by cycling from -1.4 V to -1.7 V and back vs. Pt, with no stirring in ionic liquid solution of 0.4 M TaF$_5$ and 0.4 M LiF in [BMP]Tf$_2$N).*

STEM-EDX analysis additionally revealed contaminants in the gold layer beneath the Ta coating. Figure 6 shows details of the Ta coating/Au interface. A detailed examination identified interdiffusion into Au, with Al being the predominant element, as shown in Figure 7. Aluminum content (up to 25 at. %) in the sputter-deposited gold was proved by the EDX spectrum profile analysis, shown in Figure 7. The Al diffused into the Au coating, most probably from the ionic liquid solution. The source of the Al was the glass-ceramic substrate. A negative working electrode potential, combined with a temperature of 200°C, would promote aluminum redeposition and diffusion into Au.

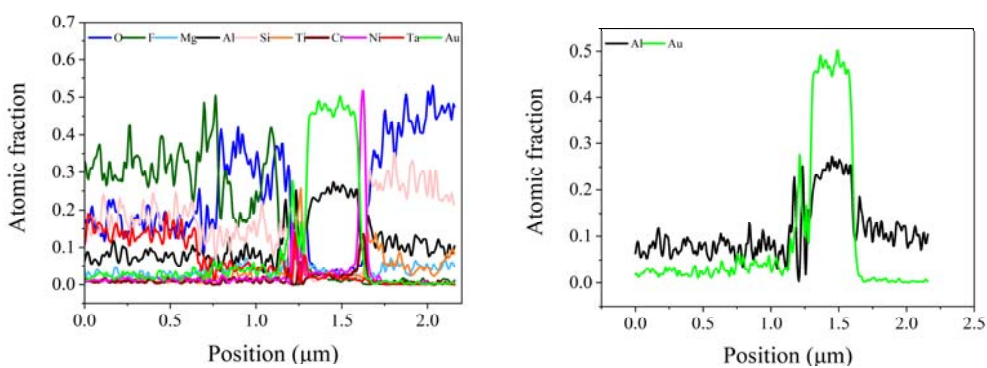

*Figure 7 EDX spectrum profile plot showing element distribution as atomic fraction across a part of Ta, Au, NiCr, and part of the glass-ceramic substrate—the same specimen as in Figure 6.*

Cavities within the Au underneath the Ta coating were revealed after the electrodeposition. Nanoporous cavities were observed before on a nitinol alloy working electrode due to the fluoride etching effect of Ni in an ionic liquid solution [18]. The diffusion of Al into the Au layer in the present study and consequent etching in the fluoride-containing ionic liquid is the most probable reason for the cavities formation. The formation of a porous Au structure is shown in Figure 8b. The Au sputter deposited on the NiCr/glass-ceramic substrate before immersion in the fluoride-containing solution is shown in Figure 8a. The Au contains several grain boundaries, which might also facilitate the interdiffusion.

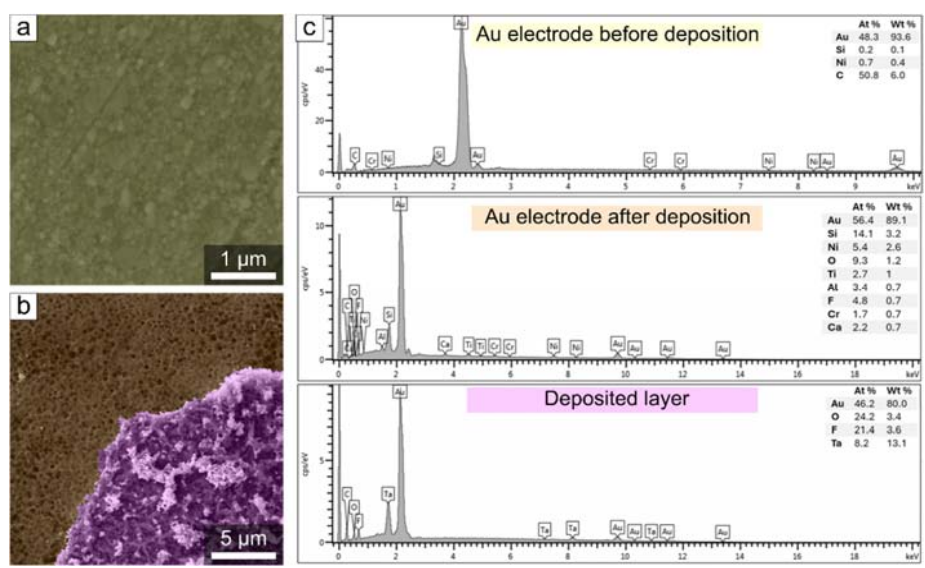

*Figure 8 SEM images of the Au-covered working electrode a) before and b) after deposition of the Ta layer in the ionic liquid. SEM images are supplemented by c) EDX analysis. The electrodeposition was performed potentiostatically at b) -1.1 V vs Pt using BMP[Tf2N] ionic liquid containing 0.25 M TaF$_5$ and 0.25 M LiF at 200°C. Before electrodeposition, cyclic voltammetry (± 1.5 V vs. Pt) and pre-conditioning (+0.5 V vs. Pt for 60 seconds) steps were performed.*

The pores were well distributed, ranging from 200 to 600 nm. Nanopores smaller than 100 nm (shown in Figure 2 in Supplementary) were obtained by increasing the fluoride concentration to 1 M and changing the working electrode potential. The nanoporous gold formation was an unexpected side result of the tantalum electrodeposition experiments.

## 4. Conclusions

Electrodeposition of tantalum (Ta) coating was performed from an ionic liquid BMP[Tf$_2$N] in the presence of dissolved TaF$_5$ and LiF at 200 °C, using two different types of working electrodes: a nanoporous alumina (PAA) template and a "planar" gold/nickel–chromium–covered glass ceramic. Ta metal coating via the PAA template with approximately 70 % Ta metal was achieved. The X-ray photoelectron spectroscopy depth analysis proved it. However, Ta metal was not detected on the glass-ceramic working electrode substrate due to high impurities, primarily oxygen, silicon, and aluminum, originating from the glass-ceramic. The wrong choice of the working electrode material (Au deposited on NiCr/glass-ceramic) led to a serious deterioration of Ta purity. Fluoride-containing ionic liquid solution in contact with the glass-ceramic material accounted for the experimental failure. Our study provided detailed analyses of the electrodeposited Ta. The elemental distribution in the thick Ta coating along its cross-sections was obtained. Nanoporous gold was achieved as a side result of the Ta electrodeposition using the Au/NiCr-covered glass-ceramic substrate. Cavities in the gold coating on the working electrode surface were observed and attributed to aluminum penetration into the gold and a consequent fluoride etching effect. Nanoporous cavities were obtained by increasing the fluoride concentration and changes in the working potential. Nanoporous gold is a highly significant material due to its unique properties, which are applicable, e.g., in catalysis and sensors. Nanoporous gold also exhibits reduced stiffness and can be engineered for specific mechanical properties, which is advantageous for applications requiring flexible materials. Additionally, the electrochemical formation of nanostructured tantalum and its oxides offers a low-cost technology for the preparation of TaO$_x$-based sensors, particularly for hydrogen gas detection.

**Acknowledgment**

The work was done under the Czech Grant Agency project no. 22-14886S. CzechNanoLab project LM2023051, funded by MEYS CR, is gratefully acknowledged for the financial support of the sample fabrication at CEITEC Nano Research Infrastructure. The authors are grateful to Dr. Alexander Mozalev of BUT for providing samples of PAA on conducting substrates with the perforated barrier layer.

**References**

[1] M. Nahra, L. Svecova, N. Sergent, and E. Chaînet, Thin tantalum film electrodeposition from an ionic liquid - Influence of substrate nature, electrolyte temperature and electrochemical parameters on deposits quality, J. Electrochem. Soc. 168 (2021) 082501. https://doi.org/10.1149/1945-7111/ac1697.

[2] P. Schütte, U. Näher, Tantalum supply from artisanal and small-scale mining: A mineral economic evaluation of coltan production and trade dynamics in Africa's Great Lakes region, Resour. Policy 69 (2020) 101896. https://doi.org/10.1016/j.resourpol.2020.101896.

[3] N.A. Mancheri, B. Sprecher, S. Deetman, S.B. Young, R. Bleischwitz, L. Dong, R. Kleijn, A. Tukker, Resilience in the tantalum supply chain, Resour. Conserv. Recycl. 129 (2018) 56-69. https://doi.org/10.1016/j.resconrec.2017.10.018.

[4] L. Xia, X. Wei, H. Wang, F. Ye, Z. Liu, Valuable metal recovery from waste tantalum capacitors via cryogenic crushing-alkaline calcination-leaching process, J. Mater. Res. Technol. 16 (2022) 1637-1646. https://doi.org/10.1016/j.jmrt.2021.12.104.

[5] G. Eranna, B. C. Joshi, D. P. Runthala, and R. P. Gupta, Oxide materials for development of integrated gas sensors - A comprehensive review, Crit. Rev. Solid State Mater. Sci. 29 (2004) 111-188. https://doi.org/10.1080/10408430490888977.

[6] S. Kim, Hydrogen gas sensors using a thin $Ta_2O_5$ dielectric film, J. Korean Phys. Soc. 65 (2014) 1749-1753. https://doi.org/10.3938/jkps.65.1749.

[7] B.-Y. Liu, W.-C. Liu, New room temperature ammonia gas sensor synthesized by a tantalum pentoxide ($Ta_2O_5$) dielectric and catalytic platinum (Pt) metals, IEEE Trans. Electron Devices 67 (2020) 2566-2572.

[8] A.P. Abbott, K.J. McKenzie, Application of ionic liquids to the electrodeposition of metals, Phys. Chem. Chem. Phys. 8 (2006) 4265-4279. https://doi.org/10.1039/B607329H.

[9] F. Endres, A.P. Abbot, D.R. MacFarlane, Electrodeposition from Ionic Liquids, Wiley-VCH, Weinheim, Germany, pp. 114-116, 2008.

[10] S.Z. El Abedin, H.K. Farag, E.M. Moustafa, U. Welz-Biermann, F. Endres, Electroreduction of tantalum fluoride in a room temperature ionic liquid at variable temperatures, Phys. Chem. Chem. Phys. 7 (2005) 2333-2339. https://doi.org/10.1039/b502789f.

[11] M. Nahra, L. Svecova, E. Chaînet, Pentavalent tantalum reduction mechanism from 1-butyl-3-methyl pyrrolidinium bis(trifluoromethylsulfonyl)imide ionic liquid, Electrochim. Acta 182 (2015) 891-899. https://doi.org/10.1016/j.electacta.2015.09.106.

[12] N. Borisenko, A. Ispas, E. Zschippang, Q.X. Liu, S.Z.E. Abedin, A. Bund, F. Endres, In situ STM and EQCM studies of tantalum electrodeposition from TaF$_5$ in the air- and water-stable ionic liquid 1-butyl-1-methylpyrrolidinium bis(trifluoromethylsulfonyl)amide, Electrochim. Acta 54 (2009) 1519-1528. https://doi.org/10.1016/j.electacta.2008.09.042.
[13] Proceeding Nanotech Tunisia 2015, (p.13). https://www.setcor.org/files/papers/1475722096_NanotechTunisia2015Proceeding.pdf (accessed 8 October 2024).
[14] A. Mozalev, J. Hubalek, On-substrate porous-anodic-alumina-assisted gold nanostructure arrays: meeting the challenges of various sizes and interfaces, Electrochim. Acta 297 (2019) 988–999. https://doi.org/10.1016/j. electacta.2018.11.192.
[15] H. Simunkova, T. Lednický, A.H. Whitehead, L. Kalina, P. Simunek, J. Hubalek, Tantalum-based nanotube arrays via porous-alumina-assisted electrodeposition from ionic liquid: Formation and electrical characterization, Appl. Surf. Sci. 548 (2021) 149264. https://doi.org/10.1016/j.apsusc.2021.149264.
[16] F. Lantelme, A. Barhoun, G. Li, J.P. Besse, Electrodeposition of Tantalum in NaCl - KCl - K$_2$TaF$_7$ Melts, J. Electrochem. Soc. 139 (1992) 1249. https://doi.org/10.1149/1.2069392.
17 P. Chamelot, P. Palau, L. Massot, A. Savall, P. Taxil, Electrochim. Acta 47 (2002) 3423. https://doi.org/10.1016/S0013-4686(02)00278-5.
[18] A. Maho, J. Delhalle, Z. Mekhalif, Study of the formation process and the characteristics of tantalum layers electrodeposited on Nitinol plates in the 1-butyl-1-methylpyrrolidinium bis(trifluoromethylsulfonyl)imide ionic liquid, Electrochim. Acta 89 (2013) 346-358. https://doi.org/10.1016/j.electacta.2012.11.026